# Quantum Interpretations

A. R. P. Rau

**Abstract**
Difficulties and discomfort with the interpretation of quantum mechanics are due to differences in language between it and classical physics. Analogies to The Special Theory of Relativity, which also required changes in the basic worldview and language of non-relativistic classical mechanics, may help in absorbing the changes called for by quantum physics. There is no need to invoke extravagances such as the many worlds interpretation or specify a central role for consciousness or neural microstructures. The simple, but basic, acceptance that what is meant by the state of a physical system is different in quantum physics from what it is in classical physics goes a long way in explaining its seeming peculiarities.

1. INTRODUCTION: GALILEAN RELATIVITY AND CLASSICAL REALITY

The state of a physical system in Newtonian mechanics is specified by the positions and velocities of its constituent masses. Newton and Galileo were already familiar with the "Galilean" Principle of Relativity, that the absolute values of these quantities are not significant, and will change from observer to observer whose own frames usually set the zero of the position and momentum scales. The laws of mechanics are invariant under such changes for all inertial frames, that is, for all observers moving with uniform velocities (including, but not exclusively, zero as in pre-Galilean conception) with respect to one another. All this is familiar even to non-physicists and high school students who have no trouble accepting that spectators in the stands will ascribe different positions to the batsman and the fielder on a playing field. Thus, one may see the motion of the ball from the bat where it was hit into the hands of the fielder who catches it as a parabolic trajectory from left to right, while someone in the opposite stands sees it as moving from right to left. That one person's left to right is another's right to left is easily accepted. Both agree on what matters, "the physics", that the ball at one moment was at the bat and at another, later, moment in the fielder's hands.

Extending to other observers who are not stationary but moving with respect to the stadium, say with constant velocity in the direction from the bat to the fielder, they will see the trajectories themselves as different. They see parabolas of varying "tightness", with the ultimate limit of straight up and down motion for the observer who moves with exactly the horizontal velocity of the ball. There is only vertical motion as seen in that frame. But all agree again on the underlying reality of the ball having been struck by the bat and ending up in the fielder's hands. While the ball always has a trajectory, not disappearing at the initial strike and spontaneously reappearing at the end but present at "all points in between", the question "what is its real trajectory?" or "what did the ball really do?" is immediately recognized as invalid or incomplete. These questions become meaningful only when they continue with "as seen by" a particular inertial frame. Only then is the question complete and legitimate. Thus, already in classical mechanics, the very recognition of the equivalence of all inertial frames serves to clarify what questions are meaningful when we try to understand the underlying reality of our physical world.

It would be possible to hold that the ball simultaneously executes all possible parabolic trajectories, any observer "choosing" one from among that infinity, a kind of "many trajectories interpretation of the motion of the ball" but most of us would see that as unnecessarily profligate simply to preserve the truncated question. Far better, following Occam, to accept that the problem lies in the framing of the question and recognize that one needs the full, completed question, the truncated version being an invalid one. Readers familiar with the many worlds interpretation of quantum physics[1] will immediately recognize its caricature here. We will return to this below.

Among the various frames is that of the ball itself (although this is a non-inertial frame, given the ball's acceleration) but the ball itself sees no motion for it! In its own, "rest", frame, it does not move; instead, it is all the other observers, the universe itself (!), which execute various parabolas. There is no surprise in that, that an infinity of others as seen by the ball execute an infinite number of trajectories. The reverse is what seems perverse, to ascribe many trajectories to the ball and have each observer *select* one from among them.

Einstein's extension of the principle of relativity beyond mechanics to electromagnetism, that is, to all of physics, clarified the basic notions of space and time. But the argument of the previous paragraphs is that already with the advent of physics in Galileo and Newton's time, it was clear that basic characteristics of our physical world, such as the equivalence of inertial observers, separate what questions can be properly posed from others, however natural they may seem, such as a small child's persistent, "But what did the ball *really* do?" *The nature of the physical world shapes the nature of our physical theories and our constructs to comprehend it*. The Special Theory of Relativity teaches us that not just the concept of a trajectory but the very concepts of spatial length and, even more startlingly, time interval, need to be seen relative to a particular observer. Two inertial observers who pass each other when both their clocks start at zero, each see the other's clock subsequently run slow. The seeming paradox is reconciled, that both are correct, by recognizing that we are talking of different times. Each, *according to his or her own time*, sees the other's clock as slow, and there is no conflict once time is recognized to be frame dependent. The "twin paradox", as usually posed, of one twin who makes an astronomical voyage and *returns* to the "stationary" (relative to an inertial observer) twin, is also resolved by recognizing the asymmetry between the two, that the one who aged less is indeed the traveller who was not in an inertial frame at all times.

## 2. TRANSFORMATION BETWEEN ALTERNATIVE REPRESENTATIONS

The recognition that basic concepts are tied to who (which frame) is specifying them introduces naturally into physics the consideration of transformations from one frame to another. "Transformation theory" plays a role analogous to that of translations between different human languages (or "representations" in physics usage). Any representation is as good as any other for describing the physics, so that in the context of Special Relativity, any inertial frame has as valid a description of the physics as any other inertial frame. No more, no less. The Lorentz transformations act as the dictionaries that relate alternative descriptions. Just as one must use some language to communicate, or even just to think about things even in one's own completely solipsistic world, one always has to use a frame to think about physical motion. But one always has to bear in mind that the "underlying reality" exists independently of any (and all) inertial frame much as the concept of redness, whether of a rose or tomato, exists independently of one's usage of 'red', 'rouge', or 'rot'.

This basic philosophy of our subject, that there exists an underlying reality of the world around us, but our models, while approaching closer and closer (we hope!) to that reality, are always separated from it by a "veil", is important to bear in mind. If nothing else, history teaches us this humility. The Newtonian worldview, which was successful for centuries, was dethroned by an Einsteinian one, and similarly a classical world by a quantum one. Our models, and the very language we use in thinking about them, inevitably use words such as position, momentum, trajectory, etc., but these are tied to our usage and who (which frame) is using them.

Quantum physics requires an even further extension of this gap (a veil or a gulf?) between the words used and the reality they aim to describe. But, qualitatively, this is no different from what has gone before as we have argued in the previous paragraphs. Our acceptance that 'position' necessarily requires specifying the frame that observes and is not a natural, stand-alone attribute of the system being observed, needs to be pushed further. It is not even an appropriate natural characteristic for a quantum system. If in some ways it is more natural to think of a classical ball in its own rest frame, a quantum hydrogen atom's natural states are the energy eigenstates of its Hamiltonian. Indeed, physics, whether of a lone hydrogen atom or anything else, can be entirely handled in the "energy representation". Of course, one can employ the "position representation" or the "momentum representation" and get as good and complete a description as in the energy representation[2].

3. ELEMENTS OF QUANTUM REALITY

Because of our greater familiarity with them, we use position or momentum to think about or work with but this is no more natural for reality as our employment of English or French or German for human communication and thinking[3]. Dirac's transformation theory[4] provides us the passage between these alternative representations of position, momentum, or energy. No physics is gained or lost by sticking to one or the other. And, a second important element of quantum physics is that the state of a quantum system may be specified either by position or momentum but not simultaneously both[5]. The uncertainty principle, that these two are mutually incompatible as expressed by their commutator being a non-zero $i\hbar$, is the primary distinction between quantum and classical physics. Indeed, all of quantum physics can be laid at the door of this commutator. It is an observation we made about a hundred years ago, that our physical universe has a non-zero value of $h$, that drives us to quantal rather than classical models to describe that world. Had this constant been zero, there would not have arisen this necessity.

Here again, it is only our bias that we usually refer to the position-momentum pair but we could equally well have picked other such conjugate pairs. While, classically, they may both be measurable in principle exactly and simultaneously, a non-zero commutator imposes an uncertainty limitation, forcing us *to pick* one or the other for alternative representations of that system. Why do we have this bias to position and momentum? It may lie in part because we are ourselves (near-) classical objects, large on the $h$ scale, so that our experiences and intuitions starting from the cradle gloss over the negligible quantal effects and make positions and velocities seem "real". And they became the basic ingredients of classical mechanics. But, here again, there is no qualitative difference from the fact that our acquaintance with non-relativistic speeds gave us little or no intuition about the seemingly strange consequences of Special Relativity. One of these is that time seemed absolute and the same for everyone but we have learnt to be comfortable with this not being true. So also must we now accept that the (quantum) world is not written in terms of position, momentum, and other such classical entities.

Once this change in ground on what is meant by the state of a system in quantum *vs* classical mechanics is grasped, the former is as deterministic as the latter. An equation of motion, the Schrödinger equation, determines exactly the state (its wave function) at a future time, little different from Newton's equations doing so similarly for a classical particle. But the distinction lies in what is meant by the state. The Hamiltonian is the operator which both provides the energy eigenstates and the time evolution. But it is not the position and momentum at any time that is determined as in classical mechanics but the wave function. And position and momentum do not commute with the Hamiltonian so that they are incompatible with energy eigenstates and time evolution. When we seek such quantities as positions and momenta of a quantal system, they are determined through averages over the wave function (all observed physics involves a product of two wave functions, whether we consider diagonal matrix elements for characteristics of a state or off-diagonal elements for transitions between states), thereby bringing in probabilities and statistics. This reflects that the underlying reality is not framed in terms of positions and momenta. Our classical bias towards positions and momenta makes us approach the physical world through those handles since we do not have apparatuses ("wave function metres") for directly sensing wave functions. It is as if we were restricted to reading a literary text always in translation or looking at the world only through spectacles.

All the puzzles and discomfort of quantum physics lie here, that in a quantum world we, and our apparatuses, are near-classical objects and thus are constrained in how we approach anything in that world. Had we the ability to probe the wave functions themselves, we would see no indeterminism. Indeed, special circumstances such as the Josephson Effect[6] or situations in the field of quantum information that sense phases permit such a direct viewing of the quantum world[7]. Otherwise, for the most part, we only see that world indirectly through position, momentum, etc.

Special Relativity already taught us that the view of objects at definite positions, and events occurring at specific instants of time, need to be replaced by a dynamics that is played out against the background of a combined space-time. Quantum mechanics goes further in saying that the underlying reality is in terms of different entities altogether, states and wave functions, but we approach or access it only in terms of position, momentum, etc.[8] Instead of invoking wave function collapse or the universe splitting with every observation (presumably at some location at some instant) into an infinitude of parallel worlds[1], it is far better to accept this limitation, that the primitives of our models do not coincide with those actual "elements of physical reality"[9]. So long as we keep the distinction in mind, we can work with either the position or the momentum representation to grasp the nature of that quantum world. Just as literary scholars with texts of a now completely extinct language and available only in translation are forever limited in grasping their exact meaning, so too will there be a gap as long as our thinking and our apparatuses are constrained by concepts natural to us such as position and momentum. But this constraint is inescapable just as for those scholars who must of course work with what they have: English, French or German. There is a *limitation* but there needs be no inconsistency in our use of any representation, say the position representation that is most common, to understand all we can of the quantum physical world[10].

4. MANY-PARTICLE ASPECTS

The idea of position becomes even more problematical when one moves beyond a single particle. The classical state of an N-particle system deals with a large set of their positions and momenta, but all these are at least vectors residing in our usual three-dimensional world. The quantum state, however, is a wave function of 3N variables, and does not sit in that world at all but in a much larger space once we go beyond a single particle. In combination with observables always involving the product of two wave functions, non-local correlations and interactions are intrinsic to many-particle quantum physics, leading to puzzles exemplified by the Einstein-Podolsky-Rosen (EPR) and Schrödinger cat constructs.

Consider first the Schrödinger cat which has figured in innumerable discussions and continues to fascinate physicists and non-physicists alike. Unfortunately, many discussions are muddled by not keeping the issues involved clear and distinct. At one level, the Schrödinger cat puzzle is only about quantum-mechanical superposition that applies also to single particles and not, as with EPR, about entanglement which needs two or more particles. Set aside first the cat itself, which to mix metaphors is a red herring. The cat serves only as an amplifier to connect from a single two-level system[11], whether an excited atom or a radioactive nucleus, to a more "lively" macroscopic biological object's two states of dead and alive.

The principal question involved in this puzzle is whether and in what sense a superposition exists. There is little doubt or ambiguity on this score. A quantum two-level system is observed only in one of two states, up or down. Upon observation, with a Stern-Gerlach apparatus or equivalent, oriented in any direction, those are the only two possible outcomes and nothing else. So also for the atom or nucleus in the box. It is only upon opening the box that one will see either the excited or the ground state. Amplified by whatever Rube Goldbergian scheme to the level of the cat, there is nothing strange then that only upon opening the box do we see whether it is dead or alive and that those are the only two possibilities. Some discussions of the poor cat being in some strange limbo between dead or alive make no sense. The premise of the set-up is that a cat is observed only in one of two states just as with a spin-1/2 or two-level atom or nucleus, and there is a one-to-one relation between the atomic states and the fate of the cat.

What then of the superposition? Again, at the level of the atom or nucleus, there is no question that superposition states exist, a hundred years of quantum studies having established that. Indeed, today, whole technologies of quantum cryptography, teleportation and computation rest on quantum superposition[12]. But these superpositions are manifest through some other measurement, not just an observation of which of two levels is realized. Either measuring with respect to some other Stern-Gerlach orientation, or monitoring an evolution in time, or comparison or coupling with some reference state is necessary. Again, setting aside the cat and considering only the microscopic atom in the box, a superposition necessarily involves also talking of the photon field. The system as a whole consists of excited atom plus no photon or decayed atom plus the emitted photon (similarly, initial radioactive isotope or residual nucleus plus alpha particle). Such a superposition has the energy in the system passing between the atom and the radiation field. Either of the subsystems by themselves oscillates between the two possibilities (excited or ground state, and zero or one photon) as a function of time. These are familiar and well observed as Rabi oscillations. Indeed, since decay can only be said to have taken place if the photon escapes to infinity, not to return to be re-absorbed (or it is somehow absorbed and removed even inside the box[13]), one can monitor happenings with a photon detector outside without having to open the box. Conversely, were the interior walls of the box to be totally reflective, the atom can continuously oscillate between excited or not, and what we have are Rabi oscillations[12]. The instant the box is opened and the photon allowed to escape is when the superposition is broken.

Extrapolating upward now to the level of the cat, the imprecision in posing the puzzle in terms of the cat becomes clear. There is no meaning specified to what a superposition means in this case or how it can be measured or monitored. Dead and alive are far from being defined even in today's biology and entirely beyond the scope of physics, whether classical or quantum. Superposition of spin amplitudes up and down has meaning but what are the amplitudes corresponding to the final outcomes of dead and alive, and what meaning is there to a superposition of such amplitudes? Whereas a photon can be reabsorbed to excite an atom, and we can even envisage an alpha particle being recaptured to make the larger nucleus, we have nothing comparable that resurrects a cat. Therefore, for all its dramatic impact, framing the puzzle in terms of the cat is meaningless. Where the problem has meaning, in terms of illuminating quantum-mechanical superposition, it is best to deal with it at the level of the atom or nucleus and there is no mystery there.

It is fairly standard nowadays to resolve the Schrödinger cat puzzle in terms of its many-particle nature and that interference between alternative states is then almost impossible to observe. Even with N of the order of 20, leave alone an Avogadro number as in a cat, energy level spacings are so small that the interference oscillations would require impossibly high time resolution for observation. Worse, the phases (as in the different time evolution of different energy states) that lead to the oscillations are easily scrambled ("decoherence"[7,12]) by the environment even with the best shielding imaginable. The universal microwave background radiation itself sets stringent constraints.

With more than one particle[11], a new quantum concept enters, namely, entanglement[12]. Stationary states of any quantum system are characterized only by the "good" quantum numbers of operators that commute with the Hamiltonian of the full system. Operators of individual particles, which do not so commute with that complete Hamiltonian, do not have definite meaning. Without going into an elaborate discussion of entanglement in this essay, we note that discussion of the EPR and other puzzles often loses sight of this, that in quantum physics subsystems cannot be said to have elements of their own reality when it is the full system that has been prepared in some fashion.

Thus, consider a state of two spin-1/2 particles. This may be in a singlet state, that is, with total spin S=0, as often used in these discussions. Once so prepared, even if the two spins spatially separate to large distances, their individual spin projections with respect to any Stern-Gerlach direction are simply not defined. This goes further than the usual statement of the paradox that if one is measured as up with respect to the z-direction, the other is necessarily down, whereas if one chooses to measure with respect to the x-direction, again if one is up the other is necessarily down. Since no spin can be simultaneously said to have definite values with respect to both z and x directions (those operators not commuting), but yet we can infer both for the distant partner depending on what we choose to measure of the one at hand, this is posed as a puzzle.

But the more important point is that no one of these individual spin projections has meaning in such a Bell set-up, only that total spin S is zero, and total $S_z$ or $S_x$ or $S_y$ are also of course zero. The issue becomes clearer were we to consider a triplet state, S=1, prepared say with $S_z$=0. This is also an entangled state and, if one spin is measured to have its $s_z$=1/2, then necessarily the other will have $s_z = -1/2$. But, now even $S_x$ or $S_y$, spin projections of the total spin, are not defined since those operators do not commute with $S_z$ and, unlike a unique value in the previous singlet case, three values of projection are now allowed for the triplet. Indeed, the prepared state does not (cannot) have zero value for these $S_x$ and $S_y$ projections but is an equal superposition of the other two values, 1 and -1. Thus, if an $s_x$ or $s_y$ on one gives either up or down, the other necessarily shares the *same* value. A system of two spins should be regarded as just that, an integral whole with the individual components subsumed.

Likewise, an N-particle system, say an N-electron atom, has a state and wave function in 3N-dimensional coordinate space and is a richly entangled state. Our description in terms of independent electron configurations, that the ground state of the helium atom is $1s^2$ or carbon $1s^2 2s^2 2p^2$, must be recognized as an approximate one. A helium atom or carbon atom born out of some chemical reaction is a state of two or six electrons with a wave function in six or eighteen dimensional space, and independent electron labels have no rigorous meaning. As well recognized in configuration interaction pictures of atoms and molecules, any state of helium or carbon is an infinite superposition of all such configurations that are compatible with the total quantum numbers, which alone are good labels[2]. Thus, in the ground state, that helium is $^1S_0$ and carbon is $^3P_0$ is all that bears meaning, these total spin S, total orbital angular momentum L, and total angular momentum J being the conserved quantum numbers (ignoring spin-orbit aspects) compatible with the full Hamiltonian. Individual electron labels and quantum numbers lose meaning and, while we often use them, it should only be with full awareness of their limitation.

## 5. CONCLUSION: THE COPERNICAN PHILOSOPHY

This essay has stressed the limitations we suffer from not just in language but in our lack of apparatuses to measure directly quantum states or wave functions and in our use of concepts such as wave, particle, position, momentum, etc., that are valid only in the classical limit of an intrinsically quantum world. Why should this limitation, that we do not speak the same language as the one in which the book of the quantal universe is written, be such a surprise? It only reinforces further the fundamental philosophical underpinning of the Copernican Principle, that there is no special status granted for any observer with regard to the physical world. Given the entire quantum universe, why should it be surprising if the physics evolved in a particular small segment of time by one biological species on an insignificant planet used concepts and words that did not coincide with the primitives of that quantal universe? To expect otherwise betrays a hubris that arises only from ignoring the distinction between the underlying reality itself and the models we make of it, even at their greatest sophistication.

**Acknowledgments**
This work has been supported by the National Science Foundation under Grant PHY 0243473 and by the Roy P. Daniels Professorship at LSU.

**Résumé**
La nature difficile de l'interprétation de la mécanique quantum vient des différences linguistiques entre elle et la physique classique. Si on posait le cas analogique de la théorie spéciale de la relativité, avec qui on avait besoin des changements de point de vue et de langage de ceux de la mécanique classique (non relative), on pourrait reconnaître les changements requis par la physique quantum. On n'a pas besoin d'invoquer des extravagances tels quels l'interprétation de plusieurs mondes ni de poser un rôle central pour la conscience ou les microstructures neuronales. Ce qu'il faut faire c'est tout simplement d'accepter la différence entre la signification d'un état de système dans la physique quantum et celle d'un état de système dans la physique classique.

involved, and practical considerations force us to truncate the expansions, so that approximations are inherent to this procedure. For long-range forces, this can even be a serious limitation in practice. Paralleling the discussions in the text, here is an instance where our limitation of not being "able to see the microscopic world directly" forces us to approach a spherically symmetric entity with macroscopic probes and detectors that lack that symmetry. This then requires extrapolations and inferences to grasp that underlying reality.

9.  A. Einstein, B. Podolsky, and N. Rosen, Phys. Rev. **47**, 777 (1935).

10.  This essay has confined itself to non-relativistic quantum mechanics, the context for most such discussions about the meaning of quantum physics. The situation with regard to position as a fundamental attribute is even more problematic at further levels of sophistication. As is well known, there is no consistent relativistic quantum mechanics of a particle. Non-relativistic quantum mechanics is at least self-consistent. But relativity and quantum mechanics cannot be melded at the level of a single particle. For the next level of consistent theories, we have to turn to relativistic field theories (currently our most successful theories). They do not ascribe a location in space for a particle. Rather, elementary particles and thereby all physical systems are viewed as excitations of underlying fields and, in principle, spread out over all space. See, for instance, J. J. Sakurai, *Advanced Quantum Mechanics* (Addison-Wesley, Reading, 1967).

11.  As in all physics, whether classical or quantal, a single particle in a potential usually means a system of two particles interacting through some physical forces and viewed in their centre of mass. Thus, a two-level atom stands for two states of a nucleus and electron with their Coulomb interaction which reduces to a single "particle" of some reduced mass moving in a Coulomb potential.

12.  M. Nielsen and I. Chuang, *Quantum Computation and Quantum Information* (Cambridge University Press, New York, 2000).

13.  Somewhat flippantly, when posed the Schrödinger cat puzzle, one should respond with "What was the color of the cat?" If a white cat, then the photon bounces around forever inside the box, the system in incessant Rabi oscillations. For an unfortunate black cat, however, when the photon hits the cat, it will be absorbed, setting off the deadly mechanism. Therefore, as surely as were the photon a bullet, the cat itself "knows" when it gets hit and is dead!